\documentclass[preprint,authoryear,12pt]{elsarticle}

\usepackage{epsfig}
\usepackage[figuresright]{rotating}
\usepackage{natbib}
\usepackage{txfonts}
\usepackage[T1]{fontenc}

\usepackage{amssymb}

\usepackage[ps2pdf,%
a4paper=true,%
breaklinks=true,%
colorlinks=true,%
pdfauthor={First Author et al.},%
pdftitle={Template for manuscripts in Advances in Space Research}%
]{hyperref}

\def\kms{km s$^{-1}$}

\journal{Advances in Space Research}

\begin{document}

\begin{frontmatter}



\title{Efficiency tests for estimating the gas and stellar
population parameters in Type 2 objects}


\author[1,2]{Bon, N.\corref{cor}}
\address[1]{Astronomical Observatory, Belgrade, 11060 Serbia}
\address[2]{Observatoire de Lyon, 9 avenue Charles Andr\'e, Saint-Genis Laval
cedex,
F-69561, France ; CNRS, UMR 5574 }
\ead{nbon@aob.bg.ac.rs}




\author[1]{Popovi\' c L. \v C.}
\author[1]{\&  Bon, E.}

\begin{abstract}
  We investigated the efficiency of estimating characteristics
of stellar populations (SP) and Active Galactic
Nuclei (AGN) emission using ULySS code. To analyse simultaneously AGN and
SP components in the integrated spectrum
of {Type 2} active galaxies, we modelled the featureless
continuum (FC) and emission lines, and we used PEGASE.HR stellar population
models provided by ULySS. In
order to validate the method, we simulated over 7000 integrated spectra of
Seyfert 2 galaxies. 
Spectra were generated using
different characteristics of the featureless AGN continuum, signal-to-noise
ratio (SNR), spectral ranges, properties of emission lines
and single stellar population (SSP) model whose initial mass function
(IMF) and abundance pattern is
similar to the solar
neighbourhood. 
Simulated spectra were fitted with ULySS to
evaluate the ability of the method to extract SP and AGN properties. We found
that the analysis with ULySS can 
efficiently restore the
characteristics of SP in spectra of Seyfert 2 AGNs, where signal-to-noise ratio
is higher than
20, and where SP contributes with more than 10\% to the total flux.
Degeneracies between AGN and SP parameters increase with increasing the
AGN continuum fraction, which 
points out the importance of simultaneous fitting of the FC and SP
contributions. 

\end{abstract}


\end{frontmatter}
\parindent=0.5 cm

\newpage
\section{INTRODUCTION}

\indent  Spectra of Seyfert 2 galaxies can {consist of} two main
components: (1) a stellar
component characterized by absorption lines from stellar atmospheres and (2) an
Active Galactic Nucleus (AGN) component {formed of} emission lines
and a featureless continuum (FC). In the
standard AGN model the UV/optical continuum is produced by non-thermal
processes, following
a power-law in the form  f$_{\lambda} \ \sim  \lambda^{\alpha}$ with a typical
value of $\alpha$ in the range $\alpha$ = [-1.5,2] \citep[e.g.][]{Kinney91,
BG92, Win92}. In the frame of
the AGN unified model \citep[see][]{Ant93}, a direct view of the continuum
source is blocked
by a dense
molecular torus {in the case of the Seyfert 2 objects}. The observed
featureless
continuum is believed to be a nuclear
starlight, scattered
by free electrons and dust. However, hot, young stars also produce a
featureless continuum that attenuates the absorption lines of the older stellar
population \citep{NW00}.
It was suggested in a number of studies \citep{Cid95, Col97, Schmitt99} that
the
UV continua in
Sy 2 galaxies may
have a considerable contribution from the nuclear star formation. Therefore, any
reliable measurement of
the emission-line spectrum of galactic nuclei has to properly account for the
influence of starlight. {On the other hand, to reliably derive the
stellar
population properties of these Sy 2 host galaxies, the contribution from the AGN
continuum and emission lines must be properly considered
\citep[see e.g.][]{Goud96, Moore02, Sarzi06}).}
{Namely, stellar absorption spectra are
contaminated by emission lines. For example, wavelength bend of Mg I lines at
$\lambda\lambda5167,5175,5184$, which are traditionally used for stellar
population studies are blended by strong emission [NI] and [FeVII]
lines.
Furthermore, absorption lines are diluted in some degree by the nuclear
continuum.\\
\indent A typical approach to investigate gas properties in the centre of an AGN
is to
{decompose the
spectrum into different components in order to remove starlight from
an integrated spectrum}. There are a number
of methods for removing the starlight contribution. The most frequently used one
is
the method of the template subtraction. 
A number of approaches have been adopted
to construct the template, as e.g.: (i)
the spectrum of an off-nuclear position within the same galaxy
\citep[e.g.][]{StorchiBergman93},
(ii) the spectrum of a different similar galaxy without emission lines
\citep[e.g.][]{HFS93}, (iii) a
mean spectrum derived from a principal-component analysis of a large set of
galaxies \citep{Hao05, VanB06} and (iv) a model spectrum constructed from
population synthesis
techniques \citep[e.g.][]{Keel83, BBA89, Sarzi05}.\\
\indent In order to minimize degeneracies between spectrum components,
it is better to avoid the subtraction of the stellar population (SP) from the
total spectrum
before the analysis of the AGN emission, and to
fit simultaneously all constituents that contribute to the total spectrum. One
of the techniques for the
analysis of an unresolved component from stellar populations in an integrated
spectrum is the spectral synthesis technique, which consists in the
decomposition of an observed spectrum in terms of
single stellar populations (SSPs) of various ages and metallicities, producing
as
an output the {star formation and chemical enrichment histories} of a
galaxy, together
with its velocity dispersion \citep[e.g.][]{Cid05,Chi07, Kol09,
MacArthur09, Cid10}. 
This is achieved by a full spectrum fitting
including the continuum 
shape and absorption features. \\
\indent One of the most popular techniques
for the spectral decomposition of AGN host galaxies is GANDALF, developed by
Marc Sarzi \citep{Sarzi06}. At the same time GANDALF fits 
velocity broadened SSP models and Gaussians representing the emission lines.
It performs the full spectral fitting of an observed spectrum with a
set of SSP models, determining the line of sight velocity distribution (LOSVD),
the best-fit stellar population model and kinematical characteristics of the
gas. The GANDALF code, however, does not consider the continuum
radiation
from the central source, which, instead, is accounted for by the method
proposed in this paper.  \\
\indent Here we present and test the method for simultaneous
analysis of gas and stellar
kinematics, star formation history and continuum radiation in the spectra of
Seyfert 2 galaxies. For
this purpose we introduce the power law continuum and emission lines in the
model of ULySS
\citep{Kol09}. The importance of
introducing the featureless AGN continuum in the analysis was explained by
\citet{Moul05}. Namely, \citet{Moul05}
showed that the SP composition is highly affected by the
presence of an additive
continuum, if this continuum is not modelled in the synthesis. \\
\indent The goal of our tests is not to model the stellar populations of
the
host galaxies, which can be complicated by the presence of star forming
regions \citep[see
e.g.][]{sm07,pop09}. On the
contrary, we study under which conditions
 an aged stellar population can
still be recovered in presence of an overlapped AGN continuum.\\
\indent The paper is organized as follows: In Section 2 and 3 we describe the
method;
in Section 4 we test its validity
and limitations, by simulating and fitting several thousands of spectra; 
in Section 5 we give the
discussion of obtained results and finally in Section 6 we outline our
conclusions.

\section{The method for stellar population fitting in AGN spectra}

\indent In order to analyse AGN spectra, we adjusted ULySS \citep{Kol09} full
spectrum fitting
package. The code was adapted to analyse
simultaneously all
components of the integrated light from an active galaxy.

ULySS\footnote{ULySS is available at: http://ulyss.univ-lyon1.fr/} generated
model
represents
bounded linear combination of non-linear components convolved with a broadening
function and
multiplied by a smooth free continuum, with a possibility to use a polynomial
additive term. The
model is generated at the same resolution and with the same sampling as the
observation and
the fit is performed in the pixel space. The fitting method consists in
minimizing the $\chi^2$ between
an observed spectrum and a model. It performs the Levenberg-Marquart
minimization \citep{Marquart63}. The code is written
in IDL/GDL starting from
PPXF\footnote{http://www-astro.physics.ox.ac.uk/~mxc/idl/ \ {The
current version of ULySS does not depend on the ppxf.}}\citep{
Cappellari04} and uses the 
MPFIT\footnote{Markwardt, http://www.physics.wisc.edu/~craigm/idl/fitting.html}
procedure.\\
\indent Single stellar population (SSP)\footnote{SSP is
population with single
age and metallicity.} models used by ULySS are spline interpolated over an
age-metallicity grid of models. {ULySS can use any library of stellar
spectra.} {Some models for stellar population spectra are provided by
ULySS
web-page, such as grid of PEGASE.HR SSPs, 
computed with the Elodie.3.1 library and a Salpeter IMF
\citep{LeBorgne04} or the grid of Vazdekis models computed with the Miles
library and Salpeter IMF (version 9, published in \citet{Vazdekis2010})}.
Therefore, by fitting a spectrum with ULySS, we reconstruct the
SSP-equivalent age and metallicity.\\
\indent ULySS has been used to study stellar populations
\citep{Bouch10,Kol11,Kol13} 
and determine atmospheric parameters
of stars \citep{Wu11}.\\
\indent Following \citet{Barth02} we defined a model M(x) of an integrated
AGN spectrum,
consisting of a stellar template spectrum T(x) convolved with a line-of-sight
velocity broadening
function G(x), a model for the AGN continuum C(x), and a sum of
Gaussian/Gauss-Hermit functions
S(x), that represent AGN emission lines:
\begin{equation}
M(x) = P(x) \{[w_0T (x) \otimes G(x)] + w_1C(x) + \sum_{i=2}^{n} w_iS_i(x) \},
\end{equation}
were P(x) is a multiplicative polynomial.\\
\indent For simplicity, we assumed a Gaussian velocity broadening function G(x),
but it is possible to
use also Gauss-Hermite polynomials (GH) of the 3rd and 4th order accounting for
deviations of the
galaxy's LOSVD from Gaussian. The 3rd
order of the GH polynomial is responsible for the asymmetry in the line profile
(h3), 
while its 4th order stands for the symmetric deviation
from a Gaussian (narrower
for positive or wider for negative h4, respectively; see e.g.
\citet{RW92,vanM94}).\\

\indent  The mismatch between the shapes of galaxy and stellar spectra occurs
because spectra of distant galaxies are obtained with different
gratings and detectors from the stellar template spectra that are used in the
spectral libraries. Therefore, the spectra of
the stars and the distant galaxies may have different continuum shapes due to
the multiplicative instrumental response function \citep[see
e.g.][]{Eriksson06}.
So, we introduced a multiplicative polynomial
P(x), into the fit to remove overall shape differences between the observed
stellar and galactic
spectra. The introduction of this polynomial in the fit
ensures that results are
insensitive to the normalization, the flux
calibration of a galaxy and stellar
template spectra, {but also to the Galactic extinction \citep{Kol08}}.
The continuum of the stellar template and galaxy
spectra is normalized during the continuum
matching process. This normalization becomes a part of the $\chi^2$
minimization, and consequently the effects
of the continuum normalization on derived velocity dispersions are, by
definition, minimized \citep{Kel00}. The
multiplicative polynomial represents a linear combination of Legendre
polynomials. \\
\indent The contribution of the components to the total flux can be calculated
from their weights ($w_i$)
which are determined at each Levenberg-Marquart iteration using a bounding value
least-square method \citep{LH95}.\\
\indent In our test we used PEGASE.HR SSP grid as it is provided by ULySS,
in the wavelength range $\lambda\lambda=[3900,6800] \ \AA$.

\section{Components of the model}

\indent Beside the stellar population, we used featureless continuum and
emission lines
in the model of integrated spectra of Seyfert 2 galaxies.

\subsection{Featureless continuum}
In the original version of ULySS, the optional additive continuum C(x) is
represented by Legendre
additive polynomial. Here we used C(x) as a non-stellar component, represented
by a power law f$_{\lambda} \ \sim  \lambda^{\alpha}$, where $\alpha$
represents the free parameter in the fit, together with the weight of this
component.\\
\indent The additive continuum was added to the stellar base to represent the
contribution of an AGN
featureless continuum (FC), a traditional ingredient in the spectral modelling
of Seyfert galaxies
since \citep{Koski78}. The spectral index $\alpha$ depends on the continuum
slope,
thus in different spectral
domains it has different values. For the optical domain, that we analysed,
the expected
value for the spectral index is usually in the interval between -1.5 and 2
\citep[e.g.][]{Kinney91,BG92,Win92}.
In the frame of the unified model \citep[e.g.][]{Ant93}, in Seyfert 2s this
FC, if present,
must be scattered radiation from the hidden type 1 Seyfert nucleus.
Nevertheless, one must keep
in mind that young starbursts can easily look like an 
AGN continuum in optical
spectra, that presents a common problem faced in spectral synthesis of Seyfert
2s \citep{Cid95, StorchiBergman00}.

The routine for the featureless continuum definition, that we use is
adequate to
the power function on the ULySS web page, tutorial 4.

\subsection{Emission lines}

When we construct a model of {narrow emission lines such as those that
can be found in Seyfert 2 galaxies}, we made a routine to define each emission
line component
separately as a Gaussian or Gauss-Hermit function. Figure
\ref{Comparison1} gives the possibility for a visual comparison 
between the real spectrum of
Seyfert 2 galaxy and the simulated one.\\
\indent Dealing with
separate emission
line models, ULySS allows
to constrain the relations between parameters of the different lines.
Theoretical
calculations \citep{Rosa85, Gal97} and measurements of the line characteristics
in the
case of ionized
nebulae and AGN \citep[see e.g.][]{Acker89, Dimitrijevic07} indicate the
relationship
between the flux of certain emission lines, as for example, the constant flux
ratio of [OIII]
4959, 5007 $\AA$ or [NII] 6548, 6583 lines. Besides, if a group of emission
lines arises in the same region,
one can expect the same widths and shifts of narrow lines. Taking all of this
into account, we set
constraints of the line properties during the fit. To constrain emission line
parameters we used TIED
parameter in PARINFO structure in MPFIT\footnote{PARINFO is an array of
structures, one for each parameter, 
used by MPFIT for simple boundary constraints which can be imposed on parameter
values. One of the entries in the 
PARINFO structure is TIED which can be used to tie one parameter value to other
parameter values.}. 
Tied parameters are considered to be fixed, and therefore no errors are computed
for them.\\
\indent Routine that we use for the emission line model construction (named
uly\underline{ }Gauss) is based on the Craig B. Markwardt's routine
gauss1.pro that could be found at:\\
http://cow.physics.wisc.edu/$\sim$craigm/idl/down/gauss1.pro.\\
\indent Similar routine that defines emission lines is provied by ULySS package
\\ (uly\underline{ }line.pro), but we did not test this routine, 
since our project started from NBurst \citep{Chi07} before the
{publication} of
ULySS.

\section{Simulations}

\indent Once we fit an AGN spectrum with the composite model, we are able to
analyse the kinematics of
the stellar component and the connection between the gas and stellar motion in a
galaxy, as well as
the age and metallicity of the stellar population.\\
\indent In order to test the method, first we performed numerical
simulations. The aims of the simulations are: (i) to evaluate the ability of
our method to disassociate the stellar population and the pure AGN spectrum from
the composite AGN + SP spectrum, which was
simulated using different fractions of the featureless continuum, and (ii) to
verify the dependencies
between the different FC contribution and characteristics of the SP spectra.\\
\indent To test the accuracy of the method, we simulated line-of-sight
integrated spectra of low luminosity AGNs. We assumed that the spectrum is
composed by the
featureless continuum, emission lines and underlying stellar population (SP)
from the host galaxy. We made a
grid of 7200 spectra with the {PEGASE.HR SSP model whose initial mass
function and abundance pattern is similar to the solar neighbourhood
(hereafter Solar-like)}, combined {with 
different} fractions and slopes of featureless continuum, different intensities
and widths of emission lines,
various spectral ranges,
SNR and degrees of Legendre multiplicative polynomial. In more details,
simulated spectra were
made with (i) AGN power-law continuum fractions in the range 10\% to 90\% to the
total continuum {(measured at 5100 $\AA$)}
and spectral indices $\alpha$=-1.5, -0.5 ,0.5 ,1.5 ,2 (ii) single
stellar
population spectra produced
with PEGASE.HR code with Solar-like characteristics (age=5 Gyr and
[Fe/H]=0
dex) and (iii)
emission lines. We convolved the SSP spectrum with the Gaussian profile
having
width (sigma) of 100 km/s in order to mimic the realistic SP velocity
dispersion for low luminosity AGNs.
We simulated spectra in three spectral ranges: (4000-6700) $\AA$, (4000-5600)
$\AA$ and (4200-5600) $\AA$, and with 10 signal-to-noise ratios (from 5 to
50). {SNR defines mean signal-to-noise ratio of the analysed spectrum.
According to this parameter, the program generates a constant error spectrum
and add it to the simulated spectrum.}\\
\indent Depending on the spectral range, we simulated emission lines (H$\delta$,
H$\gamma$, HeII 4689 $\AA$, H$\beta$, [OIII]
4959, 5007 $\AA$ lines, [NII] 6548, 6583 $\AA$ and H$\alpha$). We linked widths
of {emission} lines as
\begin{equation}
\rm \sigma_{line1}/\lambda_{line1}= \rm \sigma_{line2}/\lambda_{line2},
\end{equation} 
where $\sigma$ and $\lambda$ are widths and wavelength of corresponding line,
respectively. 
All {emission} lines shifts were fixed to the same value. Besides, we
linked intensities of the 
[OIII] lines (I([OIII]$\lambda$5007 $\AA$=3 $\times$ I([OIII]4959 $\AA$))
\citep{Dimitrijevic07}, 
as well as intensities of the [NII] lines (I([N II]6583 \AA) = 3 x I([N II]6548
\AA)).  \\
\indent Finally, varying characteristics of the line, we created 25 different
{[OIII]4959 $\AA$ }line profiles, using five different intensities and
five widths (with FWHM from
200 \kms up to 2000 \kms). {Intensities of all
the other
lines in the model were linked in some ratio with the intensity of
[OIII]4959 $\AA$ line, while their widths and shift are linked with the
{[OIII]4959 $\AA$}
line, as it is described in the previous paragraph. } \\
\indent {We made a test to analyse the dependences between the SSP 
properties and the degree of the multiplicative polynomial (md). We used 50
degrees
of the polynomial (from 1 to 50), and we found that in the case when the SP is
contributing with more than 75\% to the total
spectrum and when the SNR is higher than 20, the
restored SP properties do not depend much on the degree of the polynomial and
the result is stable for the degree of the polynomial higher than 15.
Interesting result is obtained for the SP age and metallicity.
Namely, even though the restored SP metallicities are very close to the expected
one for any md, for all md < 15 they are underestimated, and for all md > 15
they are overestimated. The result for the SP age is opposite. The
restored SP ages are overestimated for md < 15, and underestimated for md > 15.
 Since it is shown
in \citet{Kol08} that the high order terms of the multiplicative
polynomial (md > 50) are well decoupled from all the kinematical and
population parameters, the users are highly recommended to use the lowest md
that ensures accurate result. From these reasons we used md=15. \\
\indent To evaluate the ability of the method to restore characteristics of the
gas and 
stars in the nucleus and in the host galaxy, we fitted the simulated spectra
with the described models of integrated AGN spectra. {We use
uly\underline{ }ssp.pro,
uly\underline{ }power.pro and uly\underline{ }gauss.pro routines, to define the
fitting model. Therefore, we used a series of components, since each emission
line corresponds to the separate component in the fit.}\\ 
\indent In order to analyse the parametric space and dependencies between
parameters, 
we performed Monte-Carlo simulations and we made $\chi^{2}$ maps to find the
position of the global minima and possible local minima. \\

\section{Results of simulations}


To illustrate the goodness of the best fit, and its dependencies on SNR and FC
contribution, we show in
Figures \ref{PL01}-\ref{SNR35} the analysis of simulated spectra with the
integrated AGN model.
Figure \ref{PL01} represents
the fit of simulated spectra with 10\% of FC contribution and SNR=20. Panels
(a) and (b) show
the total analysed wavelength range, where one can notice that emission lines
are fitted very well,
i.e. that residuals from prominent emission lines are on the
noise level. For better
inspection of the absorption line fit, panel (c) expands a small wavelength
region,
around Mg I b $\lambda\lambda$
5167, 5173, 5184 triplet. Figure \ref{PL07} shows the best fit for the case with
70\% of
the FC contribution
to the spectrum and SNR=20. We can notice on figures that residuals are
around 3\% of the absorption signal. However, the {visual inspection} was
not sufficient for
evaluation of the dependencies between restored parameters and the FC
contribution. \\
\indent Figures \ref{SNR5} and \ref{SNR35} illustrate the differences between
the best fits of
spectra with SNR=5 and
SNR=35, respectively. In both cases the contribution of the featureless
continuum is 50\%. Residuals from the fit of the spectra with SNR=5 show that
the output is not reliable. \\
\indent Figure \ref{Age_SPfract} represents the restored SSP ages in a
response to the FC contribution to the
total spectrum (10\%-90\%) in the cases of SNR=40 (top panel) and SNR=20
(bottom panel). In both cases the method is able to restore the SP age, but if
the SP
contributes with less than 
$\sim$ 10\% to the total
spectrum, 
the uncertainty of the obtained parameters {is much higher as reflected
in the
much larger size of the error bars}. The relationship
between other SP parameters and FC contribution can be analysed from Tables
\ref{Table_SNR20}
and \ref{Table_SNR40}. These tables list
obtained results from
the single best fits in the cases of the different AGN continuum contributions
to
the
total spectrum
(10\%-90\%), and for two different signal-to-noise ratios (SNR=20 in Table
\ref{Table_SNR20}
and SNR=40 in Table \ref{Table_SNR40}). Simulated spectra were obtained for
spectral
index=1.5, I([OIII]4959)=1 and
FWHM([OIII]4959)=600 \kms. The analysed spectral range was 4000-5600 $\AA$. Our
method failed to restore kinematic properties of the
stellar population
(both the velocity dispersion
and mean stellar velocity) if the FC contribution is higher than 90\%. According
to the inspected
results, we can conclude that the adjusted ULySS provides reliable results for
the
stellar
population age, metallicity
and kinematics if SP contributes more than $\sim$10\% to the integrated AGN
spectrum.\\
\indent Figure \ref{Age_AGN_SNR} represents the response of the FC continuum
contribution and SSP age to the SNR variability. One can notice that the method
restores the FC contribution with high precision {for any of tested noise
level}. On the other
hand, {in the case of low SNR ($<$ 20)}, the restored SP age is not
reliable. In Table \ref{Table_SNRs} we give restored values for different
signal-to-noise ratios
(SNR from 5 to 50). As it can be seen from Table \ref{Table_SNRs} the  SP
metallicity is well restored for SNR higher than 10. For
lower SNR the code is
not able to restore the metallicity. Kinematics of the SP, emission line
characteristics and featureless
continuum characteristics are not that much sensitive to the noise level.\\
\indent Tables \ref{Table_4000_6700}-\ref{Table_4200_5600} show that different
spectral ranges do not affect the
sensitivity of the fitting method,
as it was expected.\\
\indent {We performed Monte Carlo (MC) simulations to visualize biases,
errors
and degeneracies between parameters, for the case when the SSP equivalent age is
5 Gyr, SSP equivalent metallicity is [Fe/H]=0 and when the featureless
continuum
fraction is 25\% and 50\%.}
MC simulations produce an array of solutions 
which can be exploited to measure the dispersion of the values of
the parameters and allow us to view the correlations. 
Figure \ref{MCFeH} illustrates the dependencies between the AGN featureless
continuum contribution and (a) mean stellar population age,
(b) mean stellar population metallicity ([Fe/H]). One can notice
that the determination of both SP characteristics depends on the FC
contribution. Moreover, the
dependences are increasing with decreasing of the stellar population
contribution
to the total continuum. {Obtained degeneracies are expected because of
the
influence of the emission lines and AGN continuum on the stellar population
spectrum. For example, SP age is mostly affected by the $H\beta$ line, so
broad emission $H\beta$ line masks the absorption one and therefore could affect
the restored parameter. Besides, FC dilutes the absorption lines such as G-band
and Mg I, that affects the SP metallicity.} \\
\indent When we analysed the dependencies between the FC fraction and
SSP age, for the case
of 50\% of the FC contribution to the spectrum (Fig. \ref{MCFeH} ), 
we noticed a small void at the expected age (5 Gyr).  We obtained similar void
even for much higher number of MC simulations. For the moment, we do not know
what caused the void, but we note that 5 Gyr is still the mean value of the
distribution of the solutions.

\section{Discussions and conclusions}

\indent {Inspections of the obtained results showed in which conditions
our
fitting
procedure is
giving highly accurate results},
for both kinematical and physical properties of stellar and gas component of an
integrated AGN
spectrum. The result depends mainly on the AGN continuum contribution to the
total
spectrum and on
the signal-to-noise ratio.\\
\indent Detailed inspection of the fit output gave parameters ranges where we
can expect to obtain
correct results for both the gas and stellar component in the spectrum, and also
gave some limitations that one has to have in mind while using this method. We
concluded that the parameters of the
gas and stars in the galactic spectrum can be well restored if the spectra have
SNR $>$ 20 and if
the stellar population fraction is higher than 10\%. In the cases of a high
noise, the restored value
of {measurement is close} to the expected one, but the error bars
increase with
the growth of
the noise. The accuracy of the fit
does not depend on the wavelengths range. Since the fit depends on the initial
values of parameters, it is very important to
analyse the parametric space before fitting. Emission lines have to be fitted
very well in order to
estimate correctly the kinematical properties of the stellar population, so one
can conclude that
the fit of stellar population can not affect the fit of emission lines, but the
fit of emission lines can
lead to the misfit of the SP. For a good fit it is very important not to misfit
some prominent {absorption} lines, such as Mg I b $\lambda\lambda$ 5167,
5173, 5184 triplet
 or G band. 
The results are almost
insensitive to the degree of multiplicative polynomial, if the SNR is
higher than 20. For
lower values of the S/N level, the fit is better for higher values of md
{($\rm md \gtrsim 15$)}.\\
\indent Monte-Carlo simulations revealed that the degeneracies between AGN and
SP parameters
increase with increasing the AGN continuum fraction. This shows that AGN
continuum and SP
spectra should be fitted at the same time. \\
\indent The test of the method leads us to conclude that characteristics of the
gas
and stars in the galaxy can be efficiently restored in
a wide range of the parameters. The tested method is mainly limited with
the
stellar population used in the model. The Solar like SSP that we used in this
work can mimic spectra of Seyfert 2
galaxies, when the observed spectra is emitted from the disk, or
from approximately the first kpc of the galaxies. 
We also tested the models for stellar populations at ages of 1, 2 and 4
Gyr, and we did not find any remarkable differences with presented results. We
note that in the case of using very young stellar population in the model, the
parametric space could be affected in a different way. Yet, an SSP  is not a
realistic spectrum, since real spectra would be more accurately represented by
linear combinations of SSPs. Therefore, a test with composite stellar
population models would be more appropriate and we are currently working on it.

\section{Acknowledgements}

This research is part of the projects 176001 ''Astrophysical
spectroscopy of extragalactic objects'' and 176003 ''Gravitation
and the large scale structure of the Universe'' supported by the Ministry of
Education
and Science of the Republic of Serbia.

We would like to thank Philippe Prugniel and Antoine Bouchard for all the help.
Also, we thank to referees for
very constructive comments that helped to improve the manuscript.




\newpage
\section{Figures}


\begin{figure}
\begin{center}
{\includegraphics[width=15cm]{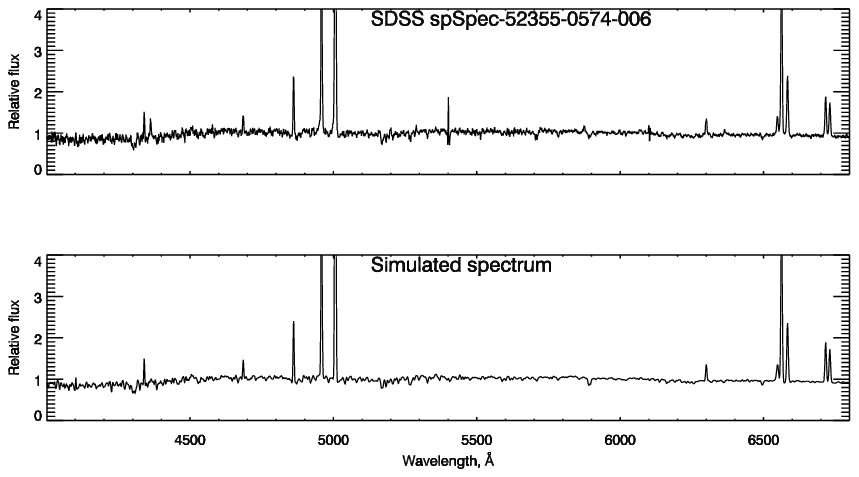}}
\caption{Comparison between real Sy2 spectrum of SDS
S galaxyspSpec-52355-0574-006 (upper plot) and
simulated one (bottom plot). } \label{Comparison1}
\end{center}
\end{figure}

  
\begin{figure}
\begin{center}
{\includegraphics[width=9cm]{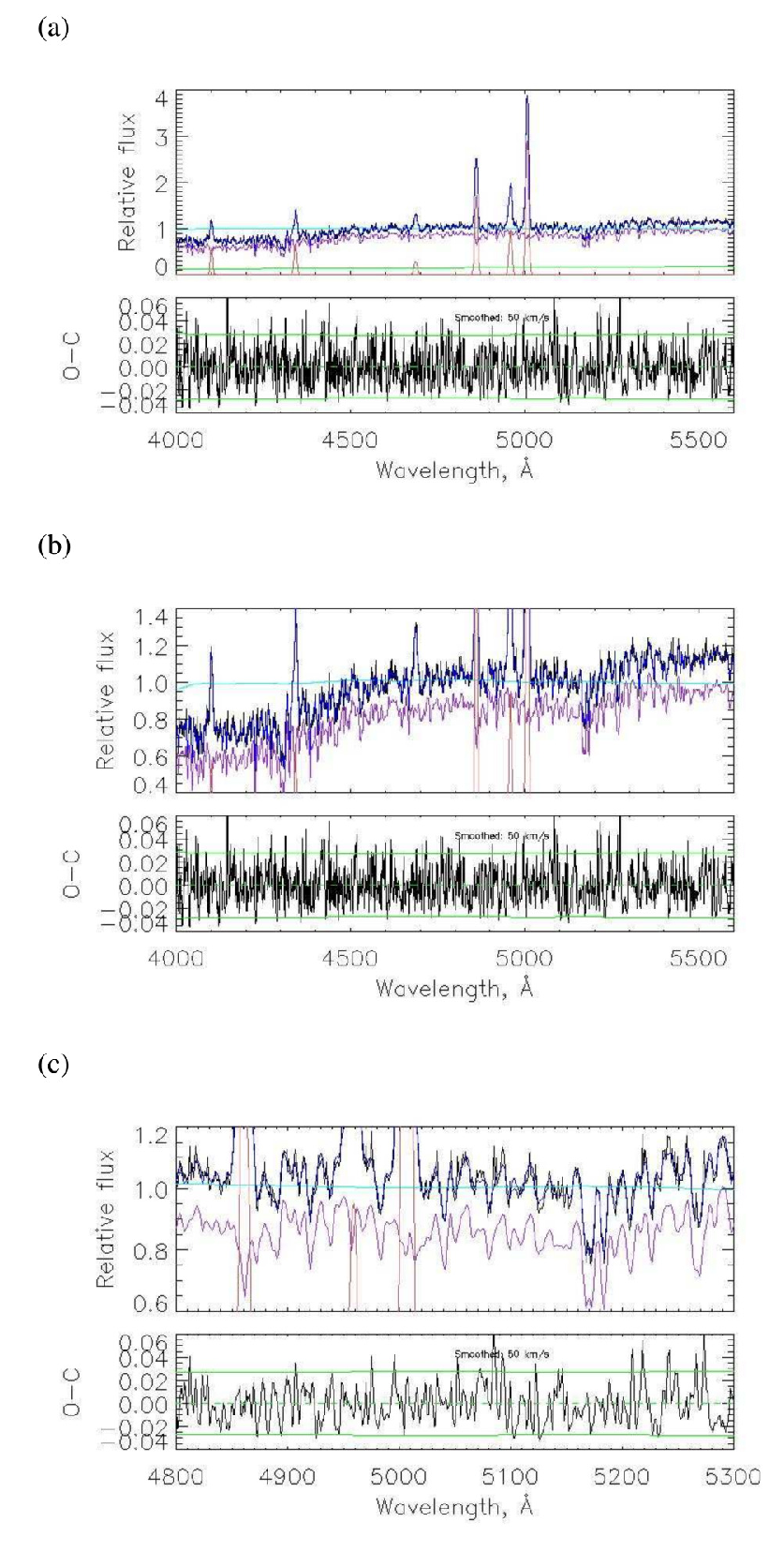}}
\caption {Panel (a) shows the fit of simulated spectra with
$\alpha$=1.5, 10\%
of the AGN contribution, intensity of [OIII]4959 equal to 1, width of 10 $\AA$,
and SNR=20. The represented wavelength range is $\lambda\lambda=[4000-5600]\AA$.
In the upper graph the black line represents the input spectrum, the
blue line
represents the best fit model, and the cyan line the multiplicative polynomial,
while the green, light red, and violet lines represent components of the best
fit model: violet -- stellar population, red-- emission lines, and green -- AGN
continuum. The bottom graph represents residuals from the best fit
(black line).
The green solid line shows the level of the noise, and the dashed line is the
zero-axis. 
In order to inspect better the fit, panels (b) and (c) expand a smaller
range of
relative flux intensity, and smaller wavelength range, respectively. }
\label{PL01}
\end{center}
\end{figure}


\begin{figure}
\begin{center}
{\includegraphics[width=9cm]{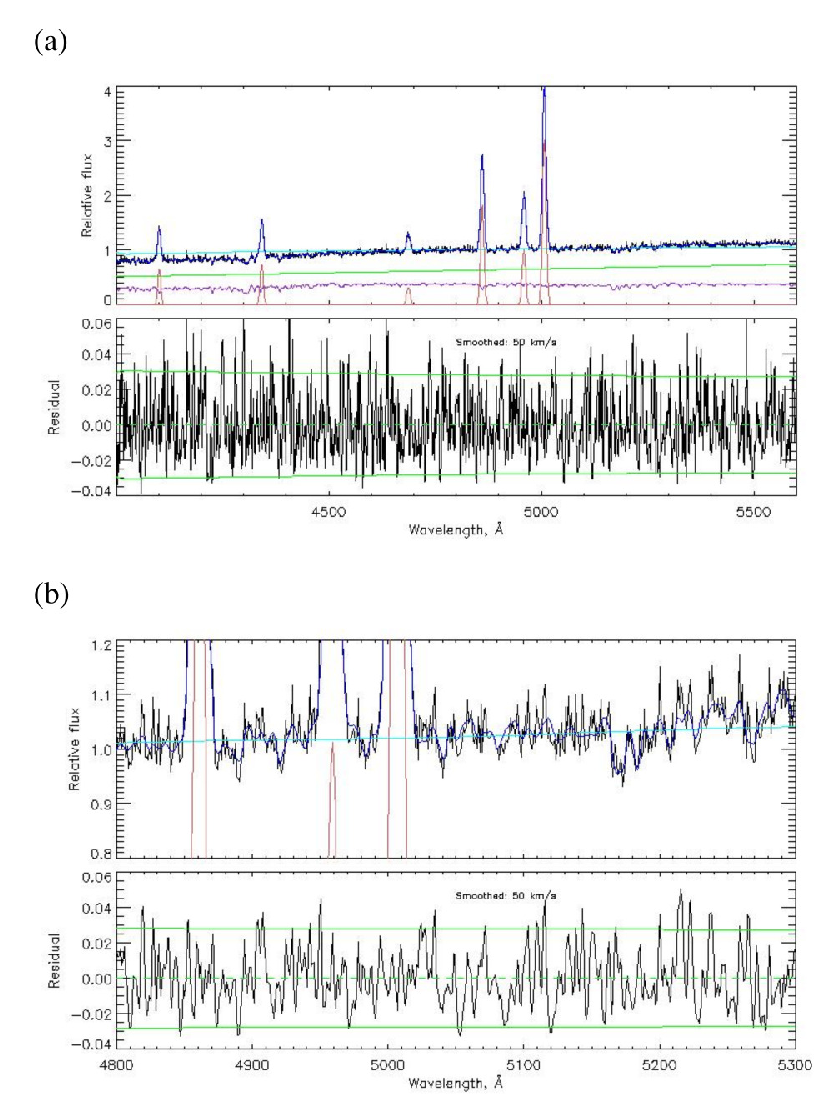}}
\caption{The panel (a) is the same as in the Fig. \ref{PL01}, but for
70\% of AGN
contribution to the total flux. The panel (b) represents the wavelength
range
between 4800 and 5300 \AA.}\label{PL07}
\end{center}
\end{figure}


\begin{figure}
\begin{center}
{\includegraphics[width=9cm]{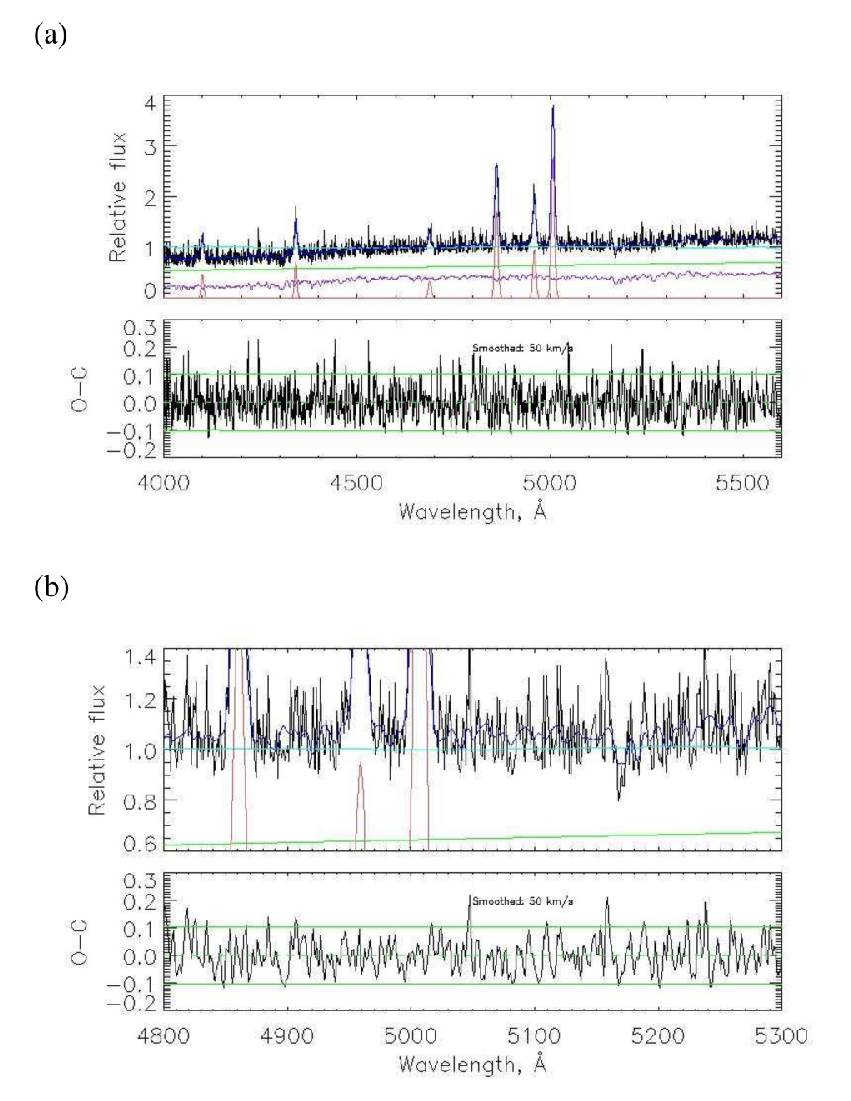}}
\caption{Fit of the simulated spectra with 50\% of the AGN contribution and
SNR=5.
The notation is the same as in the Fig. \ref{PL01}}.\label{SNR5}
\end{center}
\end{figure}


\begin{figure}
\begin{center}
{\includegraphics[width=8cm]{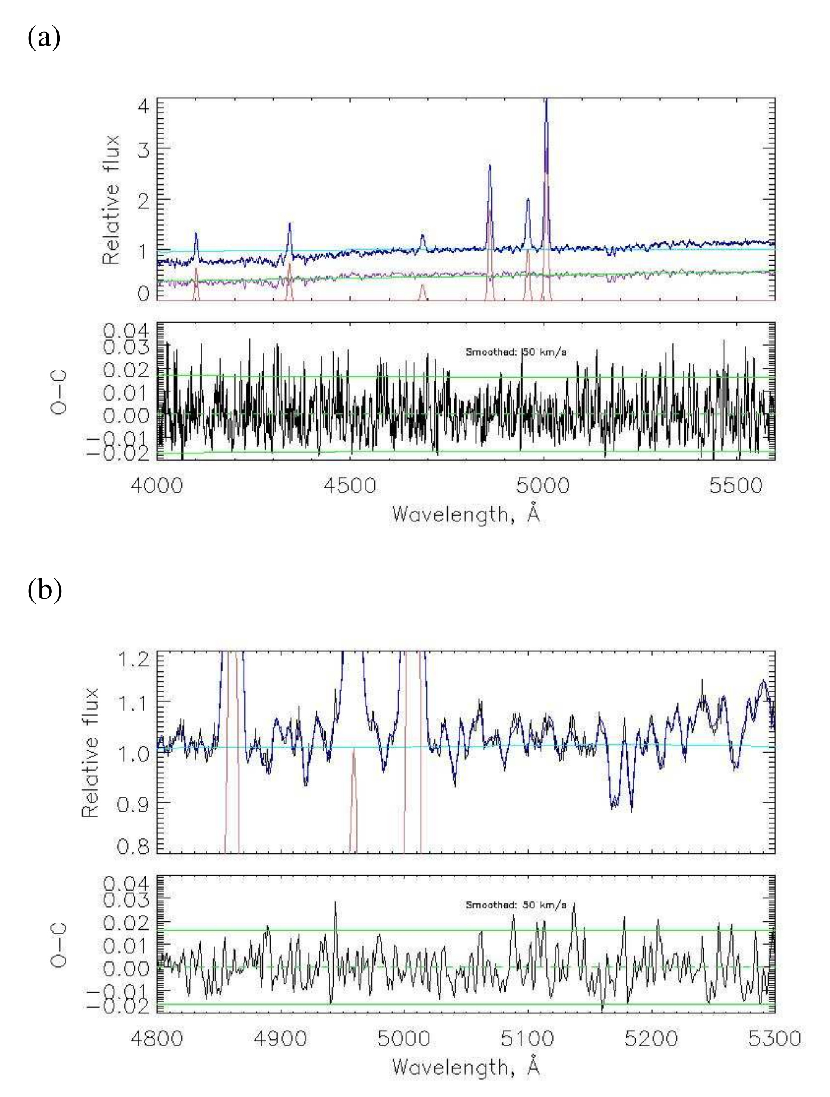}}
\caption{Fit of the simulated spectra with 50\% of the FC contribution and
SNR=35.
The notation is the same as in the Fig.\ref{PL01}}.\label{SNR35}
\end{center}
\end{figure}


\begin{figure}
\begin{center}
{\includegraphics[height=9cm,angle=270]{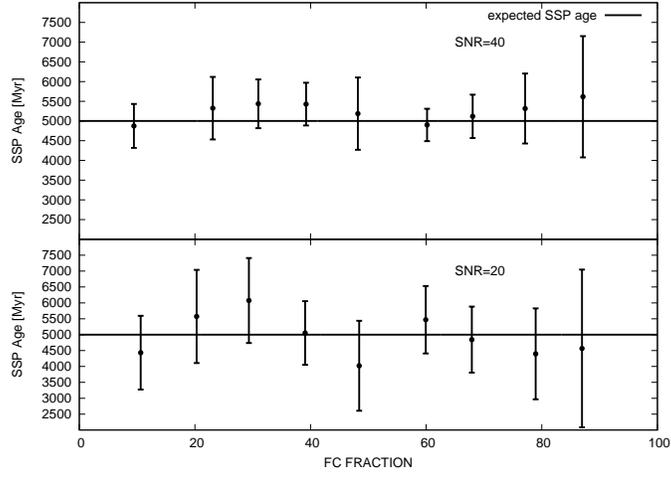}}
\caption{The restored SSP ages from the single best fit for different FC 
contributions to the total spectrum (10\%-90\%) in the cases of SNR=40 (top
panel) and SNR=20 (bottom panel). } \label{Age_SPfract}
\end{center}
\end{figure}


\begin{figure}
\begin{center}
{\includegraphics[height=9cm,angle=270]{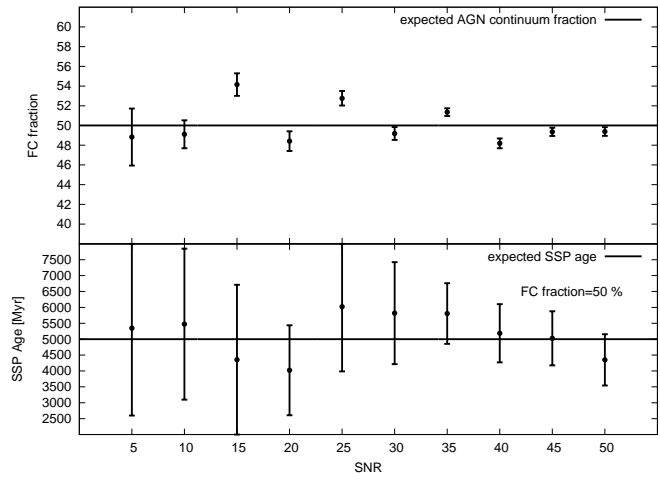}}
\caption{The restored AGN continuum fraction (top panel) and SSP age (bottom
panel) from the single best fit for different signal-to-noise ratio (SNR=5-50)
in the case of AGN continuum contribution of 50\%. } \label{Age_AGN_SNR}
\end{center}
\end{figure}



\begin{figure}
\begin{center}

(a)\hspace{.5cm}{\includegraphics[width=5.5cm]{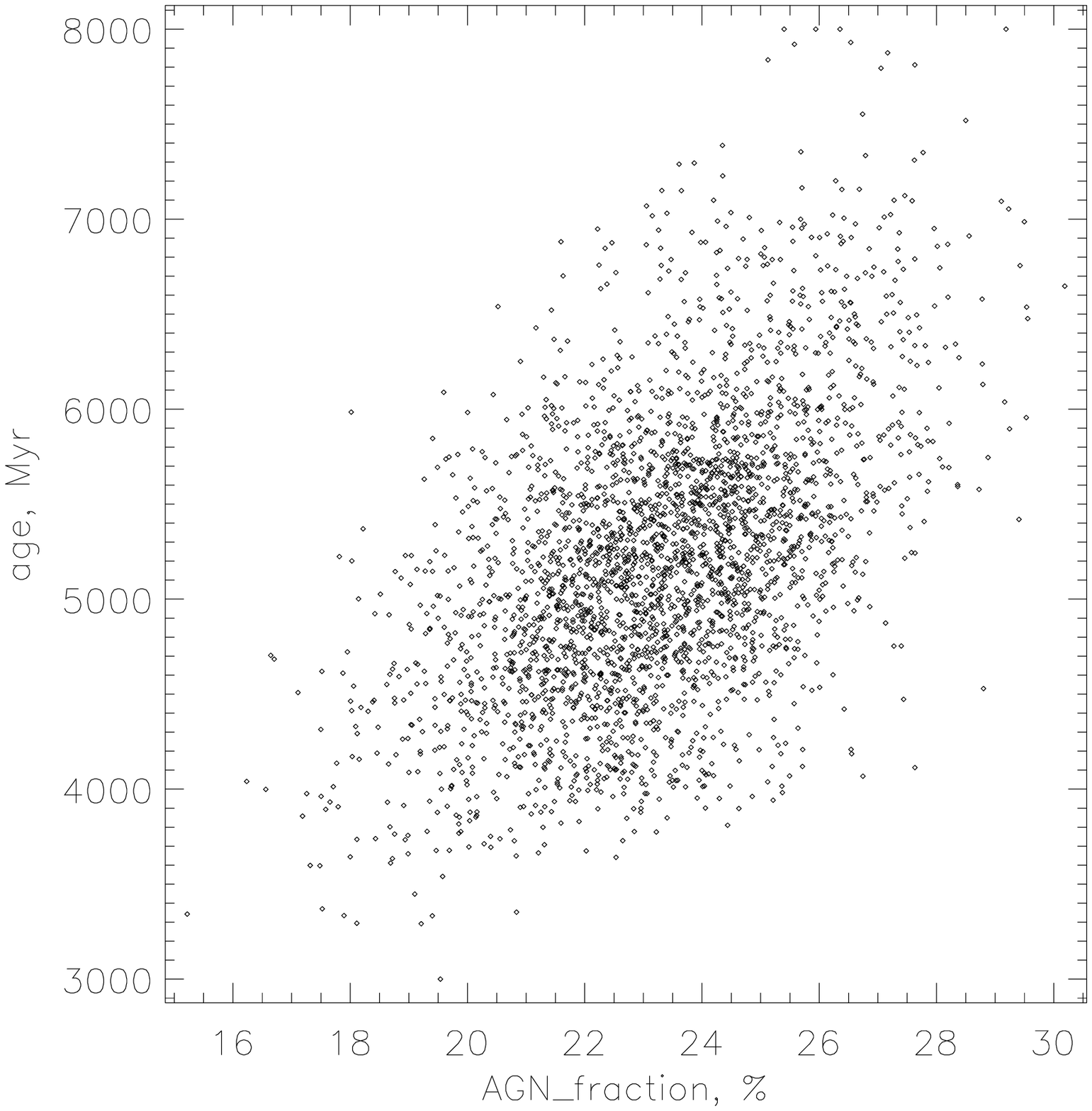}}
{\includegraphics[width=5.5cm]{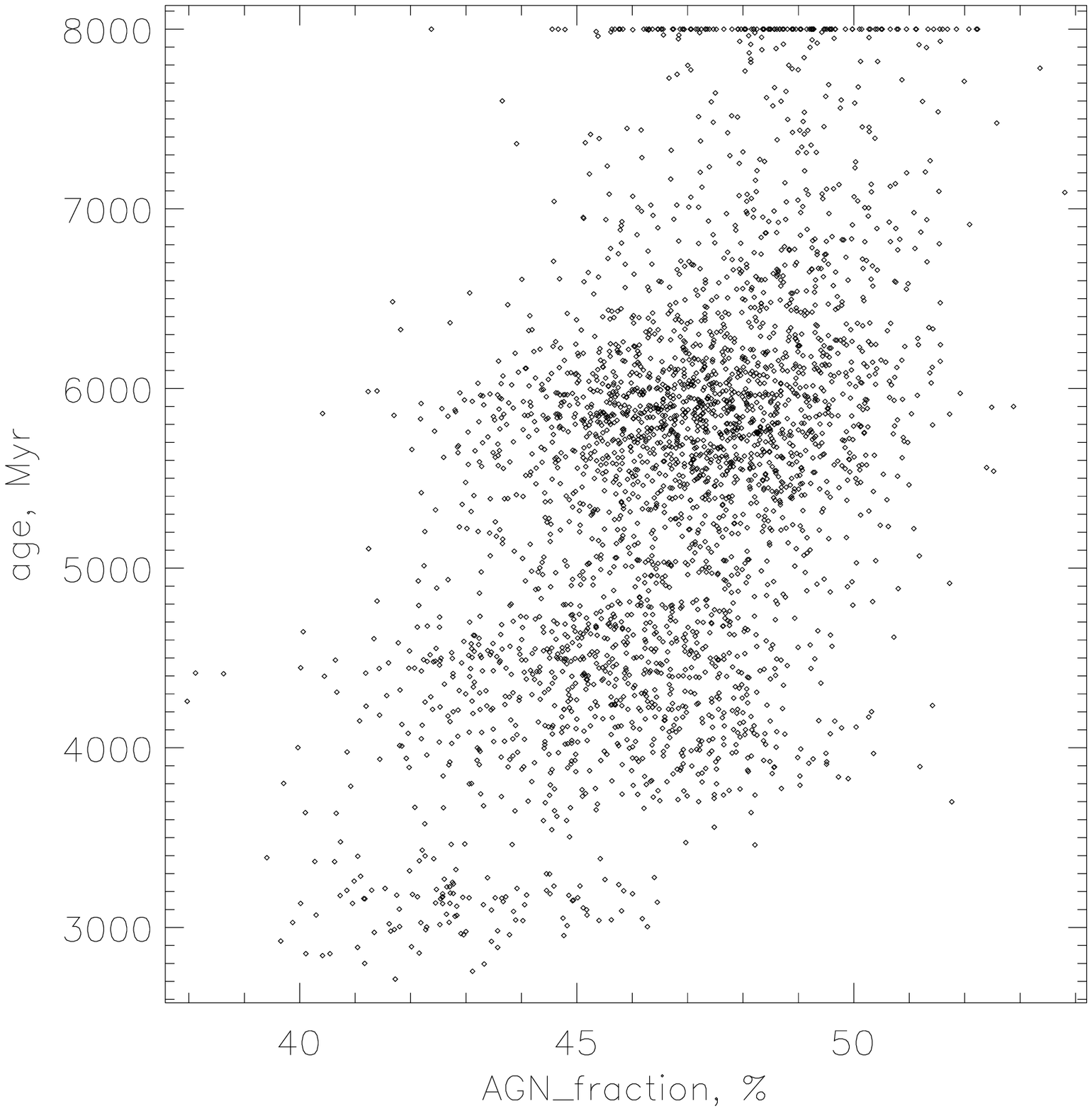}}\\
\vspace{1cm}
(b)\hspace{.5cm}{\includegraphics[width=5.5cm]{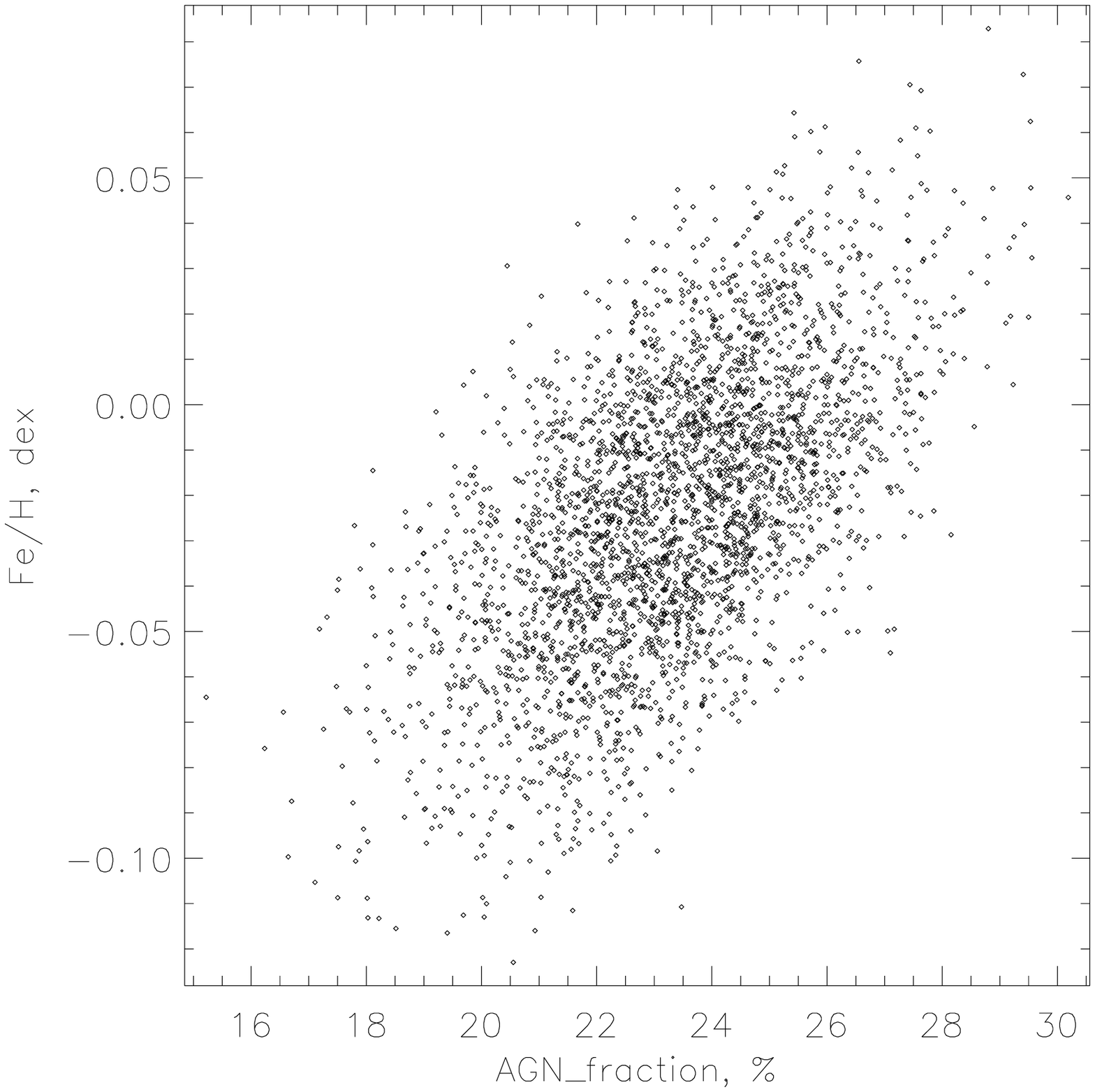}}
{\includegraphics[width=5.5cm]{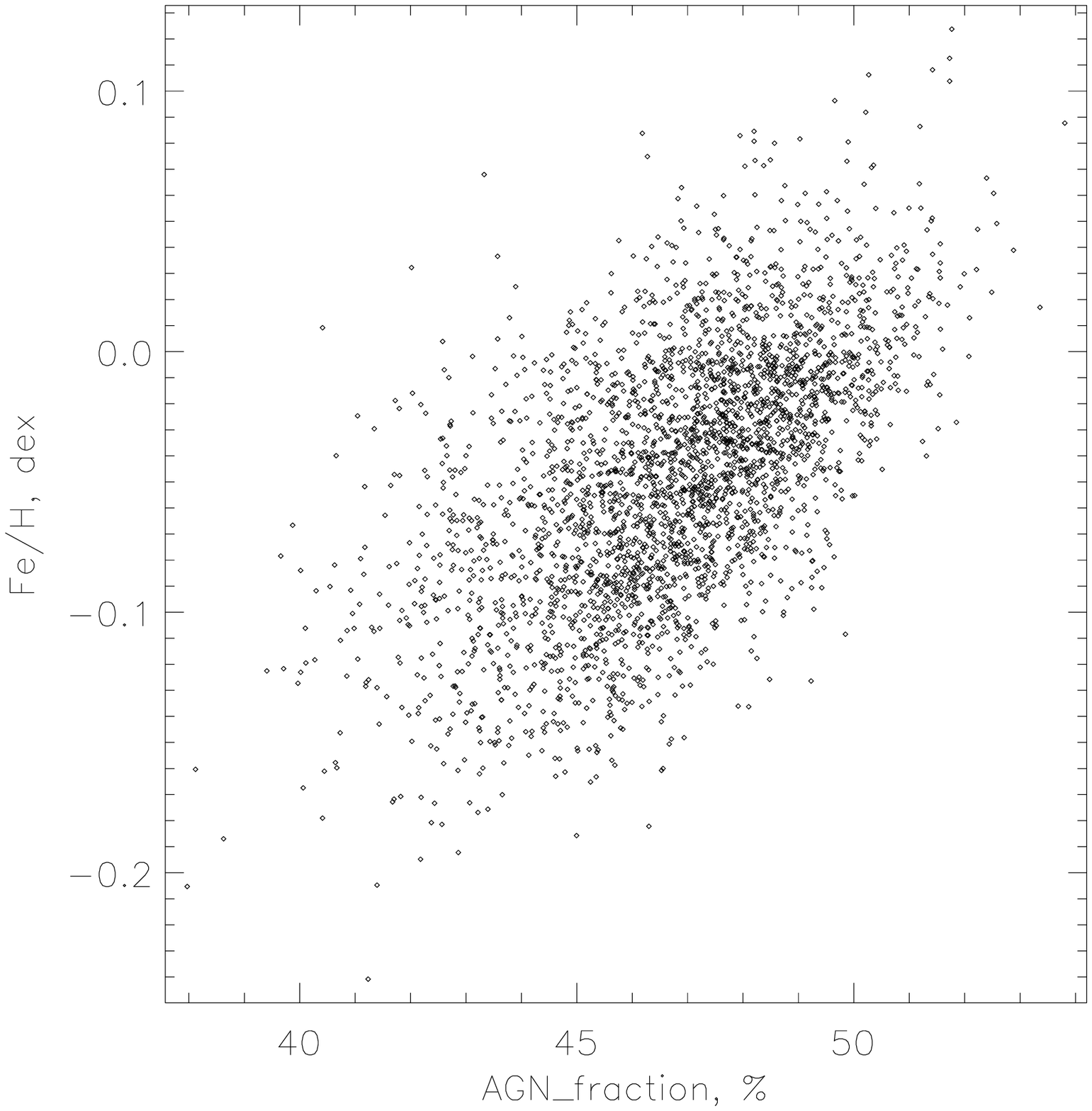}}
\caption{Result of the 3000 Monte Carlo simulations for the 
case of 75\% (left) and 50\% (right panels) of the stellar population
contribution to the continuum. Plots represent dependencies between the AGN
continuum contribution and (a) the metallicity of the dominated stellar
population (b) mean stellar population age.}\label{MCFeH}
\end{center}
\end{figure}


\begin{sidewaystable}
\begin{center}
\caption{\footnotesize Table lists initial values of parameters in the model and
obtained results from the single best fit in the cases of different AGN 
continuum contributions to the total spectrum (10\%-90\%) and a SNR=20.
Initial values in the simulation were: spectral index=1.5, I([OIII]4959 $\AA$)=1
and 
FWHM([OIII]4959 $\AA$)=600 \kms. Analysed spectral range was 4000-5600\ $\AA$.
The variables in the table are: v-mean stellar velocity, 
$\sigma_{SP}$- stellar velocity dispersion, age of the stellar population,
[Fe/H] - stellar population metalicity, $\alpha$-spectral index of the
featureless continuum, \textit{$f_{AGN}$}-restored 
fraction of the AGN continuum, \textit{$f_{SP}$}-restored fraction of the SP,
and $\sigma$-dispersion of H$\delta$, H$\gamma$, HeII 4689 $\AA$, H$\beta$,
[OIII]4959\ $\AA$ and [OIII]5007\ $\AA$ emission lines, respectively.
}\label{Table_SNR20}
\smallskip
{\tiny
\begin{tabular}{|l|r|c|c|c|c|c|c|c|c|c|}
\hline
                          &\textit{expected}    &	10\% AGN       &     
20\% AGN     &	30\% AGN      & 40\% AGN 
& 50\%AGN	          & 60\% AGN               &	70\% AGN       &     
80\% AGN     &	90\% AGN        \\
\hline
\hline
v (\kms)	          &    0	           & -0.54 $\pm$ 3.36  &  -3.09
$\pm$ 3.95 & -2.08 $\pm$ 4.13 & 4.49 $\pm$ 4.76  &	0.41 $\pm$ 6.09   & 
-11.54 $\pm$ 6.74     & -9.18 $\pm$ 8.60  & -6.84 $\pm$ 11.03 & 6.54 $\pm$ 15.47
  \\
\hline		
$\sigma_{SP}$\ (\kms)	          &   100                  & 99.84 $\pm$ 3.72  &
101.40 $\pm$ 4.47 & 97.10 $\pm$ 4.67 & 96.58 $\pm$ 5.47  & 104.30 $\pm$ 6.93    
  &  88.81 $\pm$ 7.76      & 90.69 $\pm$ 10.06 & 74.41 $\pm$ 13.45 & 67.30 $\pm$
21.49  \\		 
\hline                                                                          
                                               
Age (Myr)	          & 5000		   & 4433 $\pm$ 1160   & 5573
$\pm$ 1466   & 6075 $\pm$ 1337  & 5052 $\pm$ 1002   &   4021 $\pm$ 1415       &
5470 $\pm$ 1060        & 4844 $\pm$ 1039   & 4397 $\pm$ 1430   & 4566 $\pm$ 2481
   \\ 
\hline		
[Fe/H] (dex)	                  &   0	                   & -0.03 $\pm$
0.06  & -0.13 $\pm$ 0.08  & -0.10 $\pm$ 0.09 & -0.06 $\pm$ 0.11 
&  0.03 $\pm$ 0.09        & -0.09 $\pm$ 0.16       & -0.07 $\pm$ 0.22  & -0.12
$\pm$ 0.35  & -0.20 $\pm$ 0.64   \\  
\hline		
$\alpha$	          & 1.5                    & 1.88 $\pm$ 0.47   & 1.41
$\pm$ 0.27   & 1.65 $\pm$ 0.25  & 1.49 $\pm$ 0.19    & 1.57 $\pm$ 0.30         &
1.60 $\pm$ 0.10        & 1.54 $\pm$ 0.09   & 1.54 $\pm$ 0.08   & 1.64 $\pm$ 0.12
       \\
\hline								
$f_{AGN_{cont}}$	  &                        & 10.62 $\pm$ 0.75  & 20.30
$\pm$ 1.00  & 29.34 $\pm$ 1.04 & 39.07 $\pm$ 1.0	  & 48.41 $\pm$ 1.00    
   & 59.93 $\pm$ 1.03       & 67.90 $\pm$ 1.02  & 78.94 $\pm$ 1.10  & 86.95
$\pm$ 1.20 	\\		\hline					
$f_{SP}$	          & 	                   & 84.50 $\pm$ 0.73  & 74.85
$\pm$ 0.97  & 65.83 $\pm$ 1.02 & 56.09 $\pm$ 0.97  & 46.76 $\pm$ 0.98        &
35.26 $\pm$ 1.01       & 27.25 $\pm$ 1.01  & 16.29 $\pm$ 1.08  & 5.87e-05 $\pm$
8.4e-06 \\
\hline 								
$\sigma$(H$\delta$) 	  & 3.48                   & 3.18 $\pm$ 0.00   & 3.20
$\pm$ 0.00   & 3.22 $\pm$ 0.00  & 3.23 $\pm$ 0.00
& 3.19 $\pm$ 0.00         & 3.25 $\pm$ 0.00        & 3.26 $\pm$ 0.00   & 3.33
$\pm$ 0.00   & 3.38 $\pm$ 0.00       \\	
\hline								
$\sigma$(H$\gamma$)	  & 3.68                   & 3.37 $\pm$ 0.00   & 3.39
$\pm$ 0.00   & 3.40 $\pm$ 0.00  & 3.40 $\pm$ 0.00    & 3.38 $\pm$ 0.00	  & 3.45
$\pm$ 0.16        & 3.46 $\pm$ 0.00   & 3.53 $\pm$ 0.00   & 3.57 $\pm$ 00       
 \\		\hline		
$\sigma$(HeII 4689 $\AA$) & 4.78                   & 4.53 $\pm$ 0.42   & 4.56
$\pm$ 0.43   & 4.55 $\pm$ 0.42  & 4.38 $\pm$ 0.43
& 4.70 $\pm$ 0.41           & 4.87 $\pm$ 0.40      & 4.66 $\pm$ 0.45   & 4.79
$\pm$ 0.44   & 4.16 $\pm$ 0.38        \\
\hline										
$\sigma$(H$\beta$)	    & 4.13                 & 3.77 $\pm$ 0.04   & 3.79
$\pm$ 0.05   & 3.81 $\pm$ 0.05  & 3.80 $\pm$ 0.05
& 3.78 $\pm$ 0.06           & 3.86 $\pm$ 0.06      & 3.87 $\pm$ 0.07   & 3.95
$\pm$ 0.08   & 4.00 $\pm$ 0.11        \\
\hline 									
$\sigma$([OIII]4959\ $\AA$) & 4.21                 & 3.85 $\pm$ 0.00   & 3.87
$\pm$ 0.00   & 3.89 $\pm$ 0.00  & 3.89 $\pm$ 0.00
& 3.86 $\pm$ 0.00           & 3.94 $\pm$ 0.00      & 3.95 $\pm$ 0.00   & 4.03
$\pm$ 0.00   & 4.08 $\pm$ 0.00        \\
\hline								
$\sigma$([OIII]5007\ $\AA$) & 4.25                 & 3.89 $\pm$ 0.00   & 3.90
$\pm$ 0.00   & 3.93 $\pm$ 0.00  & 3.93 $\pm$ 0.00
& 3.89 $\pm$ 0.00           & 3.98 $\pm$ 0.00      & 3.99 $\pm$ 0.00   & 4.07
$\pm$ 0.00   & 4.12 $\pm$ 0.00        \\
\hline									
\end{tabular}
}
\end{center}
\end{sidewaystable}

\begin{sidewaystable}
\begin{center}
\caption{\footnotesize The same as in the Table \ref{Table_SNR20}, but for
SNR=40.} \label{Table_SNR40}
\smallskip
{\tiny
\begin{tabular}{|l|r|c|c|c|c|c|c|c|c|c|}
\hline
                          &\textit{expected}  &	10\% AGN        &      20\% AGN 
   &	30\% AGN      & 40\% AGN 
& 50\%AGN	          & 60\% AGN             &	70\% AGN        &     
80\% AGN     &	90\% AGN        \\
\hline
\hline
v (\kms)	          &    0	         & 0.43  $\pm$ 1.69   &  -0.57
$\pm$ 1.87 & 0.23 $\pm$ 2.13  & -0.38 $\pm$ 2.46  
& 0.13 $\pm$ 2.74         &  -3.41 $\pm$ 3.22    & -5.92 $\pm$ 4.72   & -8.28
$\pm$ 6.01  & 0.45 $\pm$ 7.74   \\
\hline		
$\sigma_{SP}$\ (\kms)	          &   100                & 100.79 $\pm$ 1.89  &
99.99 $\pm$ 2.10  & 100.4 $\pm$ 2.44 & 98.89 $\pm$ 2.83  
& 91.97 $\pm$ 3.09        &  85.94 $\pm$ 3.71    & 83.23 $\pm$        & 92.77
$\pm$ 7.20  & 74.77 $\pm$ 10.64  \\		 
\hline                                                                          
                                               Age (Myr)	          & 5000
	 & 4875 $\pm$ 558     & 5325 $\pm$ 794    & 5436 $\pm$ 618   & 5428
$\pm$ 542   
&   5187 $\pm$ 916        & 4900 $\pm$ 410       & 5118 $\pm$ 550     & 5316
$\pm$ 887    & 5615 $\pm$ 1538    \\ 
\hline		
[Fe/H] (dex)	                  &   0	                 & -0.03 $\pm$
0.03   & 0.04 $\pm$ 0.04   & 0.01 $\pm$ 0.04  & -0.04 $\pm$ 0.05 
&  -0.06 $\pm$ 0.05       & 0.00 $\pm$ 0.08      & -0.09 $\pm$ 0.11   & -0.08
$\pm$ 0.17  & 0.00 $\pm$ 0.3   \\  
\hline		
$\alpha$	          & 1.5                  & 1.28 $\pm$ 0.37    & 1.49
$\pm$ 0.15   & 1.42 $\pm$ 0.11  & 1.46 $\pm$ 0.09    & 1.51 $\pm$ 0.15         &
1.55 $\pm$ 0.40      & 1.57 $\pm$ 0.04    & 1.57 $\pm$ 0.03   & 1.63 $\pm$ 0.04 
      \\
\hline								
$f_{AGN_{cont}}$	  &                      & 9.45 $\pm$ 0.45    & 23.10
$\pm$ 0.45  & 30.96 $\pm$ 0.42 & 39.23 $\pm$ 0.48	  & 48.19 $\pm$ 0.50    
   & 60.16 $\pm$ 0.49     & 68.04 $\pm$ 0.53   & 77.12 $\pm$ 0.52  & 87.12 $\pm$
0.47 	\\		\hline					
$f_{SP}$	          & 	                 & 85.57 $\pm$ 0.44  & 71.95
$\pm$ 0.43   & 64.12 $\pm$ 0.41 & 55.85 $\pm$ 0.47  
& 46.91 $\pm$ 0.49        & 34.92 $\pm$ 0.47     & 27.04 $\pm$ 0.52  & 18.00
$\pm$ 0.51   & 7.61e-05 $\pm$ 4.39e-06 \\
\hline 								
$\sigma$(H$\delta$) 	  & 3.48                 & 3.20 $\pm$ 0.00   & 3.20
$\pm$ 0.00    & 3.20 $\pm$ 0.00  & 3.20 $\pm$ 0.00
& 3.24 $\pm$ 0.00         & 3.29 $\pm$ 0.00      & 3.29 $\pm$ 0.00   & 3.23
$\pm$ 0.00    & 3.33 $\pm$ 0.00       \\	
\hline								
$\sigma$(H$\gamma$)	  & 3.68                 & 3.39 $\pm$ 0.00   & 3.39
$\pm$ 0.00    & 3.38 $\pm$ 0.00  & 3.39 $\pm$ 0.00    & 3.43 $\pm$ 0.00	  & 3.48
$\pm$ 0.16      & 3.49 $\pm$ 0.00   & 3.42 $\pm$ 0.00    & 3.52 $\pm$ 00        
\\		\hline		
$\sigma$(HeII 4689 $\AA$)   & 4.78               & 4.56 $\pm$ 0.21   & 4.84
$\pm$ 0.23    & 4.60 $\pm$ 0.22  & 4.52 $\pm$ 0.22
& 4.41 $\pm$ 0.20           & 4.62 $\pm$ 0.22    & 4.49 $\pm$ 0.22   & 4.67
$\pm$ 0.22    & 4.80 $\pm$ 0.22        \\
\hline										
$\sigma$(H$\beta$)	    & 4.13               & 3.79 $\pm$ 0.02   & 3.79
$\pm$ 0.02    & 3.80 $\pm$ 0.02  & 3.79 $\pm$ 0.03
& 3.84 $\pm$ 0.03           & 3.90 $\pm$ 0.03    & 3.90 $\pm$ 0.03   & 3.82
$\pm$ 0.05    & 3.94 $\pm$ 0.06        \\
\hline 									
$\sigma$([OIII]4959\ $\AA$) & 4.21               & 3.87 $\pm$ 0.00   & 3.87
$\pm$ 0.00    & 3.86 $\pm$ 0.00  & 3.87 $\pm$ 0.00
& 3.91 $\pm$ 0.00           & 3.98 $\pm$ 0.00    & 3.98 $\pm$ 0.00   & 3.90
$\pm$ 0.00    & 4.02 $\pm$ 0.00        \\
\hline								
$\sigma$([OIII]5007\ $\AA$) & 4.25               & 3.91 $\pm$ 0.00   & 3.91
$\pm$ 0.00    & 3.90 $\pm$ 0.00  & 3.91 $\pm$ 0.00
& 3.95 $\pm$ 0.00           & 4.01 $\pm$ 0.00    & 4.02 $\pm$ 0.00   & 3.94
$\pm$ 0.00    & 4.06 $\pm$ 0.00        \\
\hline									
\end{tabular}
}
\end{center}
\end{sidewaystable}

\begin{sidewaystable}
\caption{The same as in the Table \ref{Table_SNR20}, but for 50\% of AGN
continuum contribution and for signal-to-noise ratios between 5 and 50.}
\label{Table_SNRs}
\smallskip
{\tiny
\begin{tabular}{|l|r|c|c|c|c|c|c|c|c|c|c|}
\hline
                          &\textit{expected}&	SNR=5       &      SNR=10      
&	SNR=15      &      SNR=20 
& SNR=25	          & SNR=30             &	SNR=35      &     
SNR=40       &	SNR=45      &      SNR=50     \\
\hline
\hline
v (\kms)	          &    0	       & -8.56  $\pm$ 13.56 & -6.80
$\pm$ 9.54  & -3.16 $\pm$ 7.04  &0.41 $\pm$ 6.09  
& -2.11 $\pm$ 4.57        &  -1.43 $\pm$ 3.68  & -1.74 $\pm$ 2.53   &  0.13
$\pm$ 2.74  & -2.43 $\pm$ 2.53  &-2.10 $\pm$ 2.56 \\
\hline		
$\sigma_{SP}$\ (\kms)	  &   100              & 82.05 $\pm$ 17.63  & 94.45
$\pm$ 10.94 & 92.69 $\pm$ 8.15  &104.30 $\pm$ 6.93  
& 91.76 $\pm$ 5.18        &  92.30 $\pm$ 4.16  & 95.09 $\pm$ 2.88   & 91.97
$\pm$ 3.09  & 100.54 $\pm$ 2.96 &96.02 $\pm$ 2.84\\
\hline                                                                          
                                               Age (Myr)	          & 5000
       & 5345 $\pm$ 2749    & 5473 $\pm$ 2375   & 4356 $\pm$ 2356   & 4021 $\pm$
1415   
&   6023 $\pm$ 2036       & 5819 $\pm$ 1606    & 5807 $\pm$ 954     & 5187 $\pm$
916    & 5026 $\pm$ 852    & 5349 $\pm$ 807\\ 
\hline		
[Fe/H] (dex)                &   0	               & 0.10 $\pm$ 0.35
   & 0.16 $\pm$ 0.19   & 0.08 $\pm$ 0.12   & 0.03 $\pm$ 0.09 
&  0.00 $\pm$ 0.08        & -0.06 $\pm$ 0.07   & 0.04 $\pm$ 0.04    & -0.06
$\pm$ 0.03  & -0.03 $\pm$ 0.05  & 0.03 $\pm$0.04 \\  
\hline		
$\alpha$	          & 1.5                & 1.51 $\pm$ 0.39    & 1.37 $\pm$
0.20   & 1.42 $\pm$ 0.44   &1.57 $\pm$ 0.30    
& 1.57 $\pm$ 0.26           & 1.53 $\pm$ 0.22    & 1.48 $\pm$ 0.09    & 1.51
$\pm$ 0.15   & 1.37 $\pm$ 0.30 &1.54 $\pm$ 0.22     \\
\hline								
$f_{AGN_{cont}}$	  &     50             & 48.83 $\pm$ 2.89   & 49.11
$\pm$ 1.42  & 54.15 $\pm$ 1.14  & 48.41 $\pm$ 1.00	  & 52.76 $\pm$ 0.74    
   & 49.19 $\pm$ 0.65   & 51.36 $\pm$ 0.39   & 48.19 $\pm$ 0.50  & 49.36 $\pm$
0.42  &49.39 $\pm$ 0.44\\ \hline					
$f_{SP}$	          & 	               & 46.89 $\pm$ 2.83  & 46.32 $\pm$
1.39   & 41.13 $\pm$ 1.11  & 46.74 $\pm$ 0.98  
& 42.38 $\pm$ 0.72        & 45.95 $\pm$ 0.63   & 43.73 $\pm$ 0.38  & 46.91 $\pm$
0.49   & 45.73 $\pm$ 0.41  & 45.67 $\pm$ 0.43\\
\hline 								
$\sigma$(H$\delta$) 	  & 3.48               & 3.21 $\pm$ 0.00   & 3.21 $\pm$
0.00    & 3.27 $\pm$ 0.00   & 3.19 $\pm$ 0.00
& 3.23 $\pm$ 0.00         & 3.24 $\pm$ 0.00    & 3.22 $\pm$ 0.00   & 3.24 $\pm$
0.00    & 3.19 $\pm$ 0.00   &3.23 $\pm$ 0.00 \\	
\hline								
$\sigma$(H$\gamma$)	  & 3.68               & 3.40 $\pm$ 0.00   & 3.40 $\pm$
0.00    & 3.46 $\pm$ 0.00   & 3.38 $\pm$ 0.00    
& 3.42 $\pm$ 0.00	  & 3.43 $\pm$ 0.16    & 3.41 $\pm$ 0.00   & 3.43 $\pm$
0.00    & 3.38 $\pm$ 00     & 3.42 $\pm$ 0.00\\		\hline		
$\sigma$(HeII 4689 $\AA$)   & 4.78             & 4.32 $\pm$ 1.56   & 3.96 $\pm$
0.69    & 5.18 $\pm$ 0.60   & 4.70 $\pm$ 0.41
& 4.78 $\pm$ 0.32           & 4.43 $\pm$ 0.27  & 4.53 $\pm$ 0.18   & 4.41 $\pm$
0.20    & 4.54 $\pm$ 0.18   & 4.60 $\pm$ 0.18 \\
\hline										
$\sigma$(H$\beta$)	    & 4.13             & 3.81 $\pm$ 0.17   & 3.81 $\pm$
0.10    & 3.87 $\pm$ 0.07   & 3.78 $\pm$ 0.06
& 3.83 $\pm$ 0.04           & 3.84 $\pm$ 0.03  & 3.81 $\pm$ 0.02   & 3.84 $\pm$
0.03    & 3.78 $\pm$ 0.03   & 3.83 $\pm$ 0.02 \\
\hline 									
$\sigma$([OIII]4959\ $\AA$) & 4.21             & 3.89 $\pm$ 0.00   & 3.89 $\pm$
0.00    & 3.95 $\pm$ 0.00   & 3.86 $\pm$ 0.00
& 3.91 $\pm$ 0.00           & 3.91 $\pm$ 0.00  & 3.89 $\pm$ 0.00   & 3.91 $\pm$
0.00    & 3.86 $\pm$ 0.00   & 3.91 $\pm$ 0.00 \\
\hline								
$\sigma$([OIII]5007\ $\AA$) & 4.25             & 3.92 $\pm$ 0.00   & 3.92 $\pm$
0.00    & 3.99 $\pm$ 0.00   & 3.89 $\pm$ 0.00
& 3.95 $\pm$ 0.00           & 3.95 $\pm$ 0.00  & 3.93 $\pm$ 0.00   & 3.95 $\pm$
0.00    & 3.90 $\pm$ 0.00   & 3.95 $\pm$ 0.00 \\
\hline									
\end{tabular}
}
\end{sidewaystable}

\begin{table}
\begin{center}
\caption{The same as in the Table \ref{Table_SNR20}, but for 50\% of AGN
continuum contribution, SNR=20 and spectral range
$\lambda\lambda=[4000-6700]\AA$.} \label{Table_4000_6700}
\smallskip
{\small
\begin{tabular}{|l|r|c|c|c|}
\hline
                          &\textit{expected}&   \textit{obtained}    \\
\hline
\hline
v (\kms)	          &    0	       & -6.54  $\pm$ 5.72  \\  
\hline		
$\sigma_{SP}$\ (\kms)	  &   100              & 92.55 $\pm$ 6.53  \\  
\hline                                                                          
                                               Age (Myr)	          & 
5000	       & 5517 $\pm$ 1596    \\   
\hline		
[Fe/H] (dex)	          &   0	               & 0.07 $\pm$ 0.09    \\ 
\hline		
$\alpha$	          &  1.5               & 1.64 $\pm$ 0.28    \\    
\hline	
$f_{AGN_{cont}}$          &   50               & 53.04 $\pm$ 0.94   \\
\hline							
$f_{SP}$	          & 	               & 40.76 $\pm$ 0.92  \\  
\hline 								
$\sigma$(H$\delta$) 	  & 3.48               & 3.24 $\pm$ 0.00   \\
\hline								
$\sigma$(H$\gamma$)	  & 3.68               & 3.42 $\pm$ 0.00   \\    
\hline
$\sigma$(HeII 4689 $\AA$)   & 4.78             & 5.06 $\pm$ 0.51   \\
\hline										
$\sigma$(H$\beta$)	    & 4.13             & 3.83 $\pm$ 0.06   \\
\hline 									
$\sigma$([OIII]4959\ $\AA$) & 4.21             & 3.88 $\pm$ 0.00   \\
\hline								
$\sigma$([OIII]5007\ $\AA$) & 4.25             & 3.95 $\pm$ 0.00   \\
\hline		
$\sigma$([NII]6548\ $\AA$)  & 5.56             & 5.21 $\pm$ 0.00   \\
\hline 									
$\sigma$(H$\alpha$)         & 5.57             & 5.17 $\pm$ 0.00   \\
\hline								
$\sigma$([NII]6583\ $\AA$)  & 5.59             & 5.24 $\pm$ 0.00   \\
\hline										
	
\end{tabular}
}
\end{center}
\end{table}

\begin{table}
\begin{center}
\caption{The same as in the Table \ref{Table_4000_6700}, but for spectral range 
$\lambda\lambda=[4000-5600]\AA$. } \label{Table_4000_5600}
\smallskip
{\small
\begin{tabular}{|l|r|c|c|c|}
\hline
                          &\textit{expected}&	\textit{obtained}      \\
\hline
\hline
v (\kms)	          &    0	       &        0.41  $\pm$ 6.09  \\  
\hline		
$\sigma_{SP}$\ (\kms)	  &   100              &        104.30 $\pm$ 6.93  \\  
\hline                                                                          
                                               Age (Myr)	          & 
5000	       &        4020 $\pm$  1415   \\   
\hline		
[Fe/H] (dex)	          &   0	               &        0.03 $\pm$ 0.09 
  \\ 
\hline		
$\alpha$	          &  1.5               &        1.57 $\pm$ 0.30    \\  
 
\hline	
$f_{AGN_{cont}}$          &   50               &        48.41 $\pm$ 1.00   \\
\hline							
$f_{SP}$	          & 	               &        46.74 $\pm$ 0.98  \\  
\hline 								
$\sigma$(H$\delta$) 	  & 3.48               &        3.19 $\pm$ 0.00   \\
\hline						 		
$\sigma$(H$\gamma$)	  & 3.68               &        3.38 $\pm$ 0.00   \\    
\hline
$\sigma$(HeII 4689 $\AA$)   & 4.78             &        4.70 $\pm$ 0.41   \\
\hline										
$\sigma$(H$\beta$)	    & 4.13             &        3.78 $\pm$ 0.06   \\
\hline 									
$\sigma$([OIII]4959\ $\AA$) & 4.21             &        3.86 $\pm$ 0.00   \\
\hline								
$\sigma$([OIII]5007\ $\AA$) & 4.25             &        3.89 $\pm$ 0.00   \\
\hline		
\end{tabular}
}
\end{center}
\end{table}

\begin{table}
\begin{center}
\caption{The same as in the Table \ref{Table_4000_6700}, but for spectral range 
$\lambda\lambda=[4200-5600]\AA$.} \label{Table_4200_5600}
\smallskip
{\small
\begin{tabular}{|l|r|c|c|c|}
\hline
                          &\textit{expected}&   \textit{obtained}       \\
\hline
\hline
v (\kms)	          &    0	       & -2.91  $\pm$ 6.33  \\  
\hline		
$\sigma_{SP}$\ (\kms)	  &   100              & 97.07 $\pm$ 7.34  \\  
\hline                                                                          
                                               Age (Myr)	          & 
5000	       & 4628 $\pm$  1117   \\   
\hline		
[Fe/H] (dex)                &   0	               & 0.02 $\pm$ 0.12
   \\ 
\hline		
$\alpha$	          &  1.5               & 1.63 $\pm$ 0.52    \\    
\hline	
$f_{AGN_{cont}}$          &   50               & 49.75 $\pm$ 1.25   \\
\hline							
$f_{SP}$	          & 	               & 45.32 $\pm$ 1.24  \\  
\hline 								
$\sigma$(H$\gamma$)	  & 3.68               & 3.38 $\pm$ 0.00   \\    
\hline
$\sigma$(HeII 4689 $\AA$)   & 4.78             & 4.40 $\pm$ 0.42   \\
\hline										
$\sigma$(H$\beta$)	    & 4.13             & 3.79 $\pm$ 0.06   \\
\hline 									
$\sigma$([OIII]4959\ $\AA$) & 4.21             & 3.86 $\pm$ 0.00   \\
\hline								
$\sigma$([OIII]5007\ $\AA$) & 4.25             & 3.90 $\pm$ 0.00   \\
\hline		
\end{tabular}
}
\end{center}
\end{table}

\end{document}